\documentclass[fleqn,10pt]{wlscirep}
\usepackage[utf8]{inputenc}
\usepackage[T1]{fontenc}
\title{Experimental determination of line energies, line widths and relative transition probabilities of the Gadolinium L X-ray emission spectrum}

\author[1,*]{Malte Wansleben}
\author[1]{Yves Kayser}
\author[1]{Philipp Hönicke}
\author[1]{Ina Holfelder}
\author[1]{André Wählisch}
\author[1]{Rainer Unterumsberger}
\author[1]{Burkhard Beckhoff}

\affil[1]{Physikalisch-Technische Bundesanstalt (PTB), Abbestraße 2-12, 10587 Berlin, Germany}

\affil[*]{malte.wansleben@ptb.de}

\begin{abstract}
In this work the complete L-emission spectrum of gadolinium with respect to line energies, natural line widths, and relative transition probabilities was investigated using monochromatized synchrotron radiation. The measurements were realized in the PTB laboratory at BESSY II by means of an in-house built von Hamos spectrometer based on up to two full-cylinder HAPG mosaic crystal. The von Hamos spectrometer is calibrated by means of elastically scattered photons from the employed synchrotron radiation beamline leading to a well-defined energy scale and an experimentally determined spectrometer response. A selective excitation of the gadolinium L subshells was carried out to ensure a robust deconvolution of neighboring emission lines of different L subshells. The experimental results are discussed in the context of existing data from common databases and published values since significant deviations, especially for the L$\gamma_2$ and L$\gamma_3$ emission lines, are observed. We further substantiate and discuss two satellite lines at the low-energy side of the L\(\beta_{2,15}\) and L\(\gamma_1\) emission lines arising from the N\(_{4,5}\) subshell.
\end{abstract}
\begin{document}

\flushbottom
\maketitle
%
%
%
\thispagestyle{empty}


\section{Introduction}

Qualitative and quantitative characterization of increasingly complex material systems is a challenge in industrial and scientific applications. In many cases non-destructive and non-preparative methods, such as X-ray fluorescence (XRF) analysis, represent a very elegant way to study these systems. In XRF either particles or X-rays serve to excite the atoms in the specimen and subsequently emitted element specific fluorescence radiation is detected and analyzed. As such, XRF represents a reliable tool in quantifying the elemental composition of a specimen provided that suitable reference specimen are available. Reference-free XRF\cite{Beckhoff2008a} is based on atomic fundamental parameters and calibrated instrumentation and enables thus a quantitative analysis independent of reference materials or well-known specimen. The reliability of this method, however, is highly affected by the accuracy on the knowledge of the fundamental parameters of the chemical elements to be analyzed.  

The Physikalisch-Technische Bundesanstalt (PTB), Germany’s national metrology institute, determines atomic fundamental parameters by means of the reference-free XRF approach \cite{Honicke2014,Menesguen2015,Honicke2016,Unterumsberger2018}. In X-ray spectrometry the term fundamental parameters (FP) generally refers to element specific parameters relevant to the production of XRF radiation such as 
fluorescence yields, line widths and energies, relative transition probabilities and, in the case of L-shells and higher shells, Coster-Kronig transition probabilities. 
In this work, we address the line energies, line widths and relative transition probabilities for the L-subshell lines of gadolinium (Gd) using a wavelength-dispersive von Hamos X-ray spectrometer\cite{Holfelder2018}. The determination of the relative transition probabilities is supported by a calibrated energy-dispersive silicon drift detector (SDD).

The chemical element of Gd belongs to the group of rare-earth metals (REM) and offers a richness of specialized uses from neutron therapy for tumor treatment\cite{Tokumitsu2000}, contrast agent for medical MRI\cite{Caravan1999}, to shielding in nuclear reactors. Gd has a half-filled 4f shell leading to a fairly large magnetic momentum per atom. Among other properties, this makes Gd an ideal system to study magnetic ordering in terms of charge and spin dynamics \cite{Carley2012}. While K emission of REMs demands X-ray sources delivering photon energies of 40 keV and upward, the L emission is energetically comparable to K emission of 3d transition metals and, hence, can be accessed more conveniently in terms of excitation and detection systems. Consequently, detailed knowledge on the L emission of REMs is very much desired. 
A recent publication addressed the reassessment of the absolute line energies of L emission lines of various REM using a cryogenic transition-edge sensor\cite{Fowler2017}. The authors noted various discrepancies to existing databases encouraging follow-up experiments on this matter, adding additional motivation and interest in investigating the L-emission of Gd. 

\section{Instrumentation}

The experiments were carried out at the four-crystal-monochromator beamline\cite{Krumrey1998} in the PTB laboratory at the electron storage ring BESSY II. The bending magnet beamline provides monochromatized synchrotron radiation between 1.75 keV and 10.5 keV by means of four Si(111) or InSb(111) crystals. For the presented emission experiments, the beamline was operated in the range of 5.5 keV to 9.5 keV spanning an energy range across the ionization thresholds of the three L-subshells and providing on average $4.5\times10^8$ photons/s. The beamline resolution in this regime is about \(\Delta E/E\) = 10\(^{-4}\) which has been experimentally determined by Krumrey and Ulm\cite{Krumrey2001}. A transmission diode in front of the measurement chamber allows for an accurate monitoring of the relative beamline flux during measurements. A calibrated photodiode is used to determine the incident photon flux using the experimentally determined responsivity\cite{Gerlach2008}. 
The excitation energies were tuned to energies between the L absorption edges (L-edges) of Gd to achieve a selective excitation of the different L-subshells. This allows for a straightforward assignment of the emission lines to the respective subshell as well as an improved fitting procedure of the spectra in the case of overlapping peaks from different subshells as the emission lines of the respective subshell emerge accordingly\cite{Kolbe2012}.

\begin{figure}
	\begin{center}
	 \def\svgwidth{1.0\columnwidth} 
	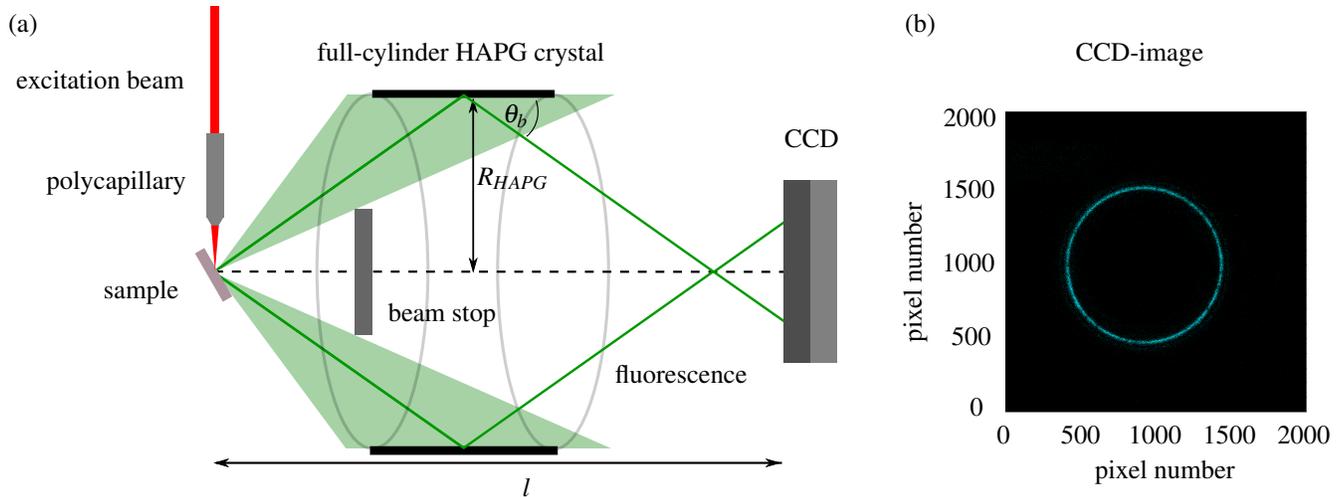
	\end{center}
	\caption{(a) Schematic topview of the full-cylinder von Hamos spectrometer and (b) summation of 150 CCD-images of the Gd L\(\alpha_{1}\) and L\(\alpha_{2}\) emission lines.}
		\label{pic:von_Hamos}
\end{figure}

We used a compact von Hamos spectrometer based on a full-cylinder crystal\cite{Holfelder2018}. A schematic of the spectrometer is depicted in Figure \ref{pic:von_Hamos} (a). The design of the instrument is similar to the prototype von Hamos spectrometer presented in Anklamm et al.\cite{Anklamm2014} and Malzer et al. \cite{Malzer2018} but with a significantly reduced radius of curvature. Furthermore, a double crystal concept is incorporated in our setup allowing for an increased resolving power up to $E/\Delta E$=2700 as it was demonstrated for titanium K\(\beta\) emission\cite{Wansleben2018}. In this work, however, only one crystal was used in order to maximize the energy band width and detection efficiency of the spectrometer. The Bragg crystal is a 40 \(\mu\)m thick highly annealed pyrolytic graphite (HAPG) sheet mounted on a full-cylinder Zerodur substrate. HAPG has been subject to detailed characterizations with respect to its use as dispersive elements \cite{Gerlach2015} as well as its diffraction properties \cite{Schlesiger2017}. HAPG is a modified material similar to highly oriented pyrolytic graphite which has been proven to be a promising material for X-ray optics\cite{Chabot1991,Arkadiev2007,Doppner2008,Grigorieva2019}. The full-cylinder maximizes the solid angle of detection and, hence, the detection efficiency. The radius of curvature of the cylinder of just 50 mm results in a very compact device which is ideal for the flexible use at different beamlines or X-ray sources. While having a small radius, the cylinder has a length of 20 mm covering a wide range of Bragg angles at a fixed position. This enables an accessible energy range which is relatively large compared to other wavelength-dispersive spectrometer designs. Due to the full-cylinder geometry the surface normal to the detector, a back-illuminated thermoelectrically cooled in-vacuum CCD (2048 pixels \(\times\) 2048 pixels, 13.5 \(\mu\)m \(\times\) 13.5 \(\mu\)m pixel size), is placed parallel to the cylinder axis. This leads to the detection of circles due to the rotational symmetry of the cylinder (cf. Figure \ref{pic:von_Hamos} (b)). A polar integration provides the final spectrum. Throughout the experiment, the detector was placed behind the respective focus position. This results in an energy dependency of the detected circle radius where the detected photon energy decreases with increasing radius. The Gd spectra were obtained by summing over 150 CCD-images at each incident photon energy and each spectrometer setting used. Each image is recorded with an exposure time of 20 s and corrected by a dark frame of the same exposure time to account for unwanted background light and thermal noise. 


The performance of the present von Hamos spectrometer in terms of resolving power is highly affected by the finite volume from which the fluorescence radiation originates. A polycapillary half-lens with a working distance of 3.5 mm is introduced to the excitation beam path to focus the synchrotron radiation onto the specimen. Polycapillary optics are widely used for all types of X-ray analysis\cite{Smolek2010,Kayser2014}. 
The alignment of the polycapillary is achieved with a five-axis piezo manipulator enabling two rotational and three translational degrees of freedom. 
A horizontally placed razor blade was used to determine the vertical dimension of the spot size. 
The resulting diameter of the spot size produced by the focusing polycapillary half-lens is determined by the first derivative of the knife-edge scan. The derivative is then fitted using a Gaussian and the full width at half maximum (FWHM) defines the spot size\cite{Unterumsberger2012}. Applying this procedure and assuming a symmetrical spot, sizes of around 60 \(\mu\)m were reproducibly attained during the course of the experiments. 
Furthermore, a photodiode on the sample holder allowed for a measurement of the transmission through the focusing capillary in order to determine the total flux impinging on the sample. 
The sample was Gd layer with a nominal thickness of 250 nm deposited on a 500 nm silicon nitride window and covered by 40 nm aluminum (Al) on top as cap layer to prevent oxidation of the Gd layer. A 4 \(\mu\)m unsupported gold (Au) foil was used as a scattering target to calibrate the energy scale and determine the response of the von Hamos spectrometer by means of elastic scattering of monochromatized photons with known energy. The advantage of Au over lighter elements is its high elastic scattering coefficient, a high ratio of elastic to inelastic scattering coefficient, as well as no fluorescence lines in the experimentally probed energy range. This ensures a high signal-to-noise ratio as well as one distinct scattering peak in the measured spectra. All samples were aligned with an angle of incidence of 45\(^{\circ}\).

\section{Calibration \label{sec:calibration}}

The Gd L-emission spreads over an energy range of roughly 2.5 keV. Since the energy bandwidth of the used von Hamos spectrometer covers more than 600 eV in the examined spectral range, only four different spectrometer settings were needed to cover the entire emission range. Each setting required an independent calibration of the spectrometer with respect to energy scale and spectrometer response. 
\begin{figure}
	\begin{center}
	 \def\svgwidth{1.0\columnwidth} 
	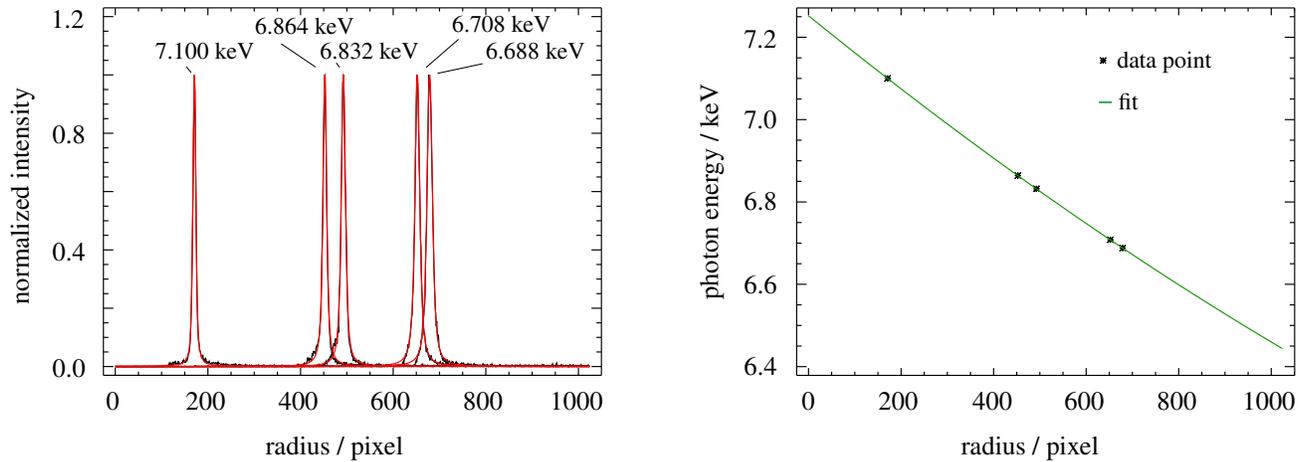
	\end{center}
	\caption{Exemplary energy calibration for one of the four von Hamos spectrometer settings. The beamline is tuned to expected positions of the respective fluorescence line energies and the signal from the elastic scattered photons is measured (left) in order to relate the radius channels of the CCD to photon energies (right). The uncertainties of the energy and the radius (0.5 pixel) are too small to be visualized.}
		\label{pic:Gd_energy_axis_pos2}
\end{figure}
A commonly used procedure to calibrate crystal spectrometers is based on measuring known standard samples and refer the energetic position of the measured emission lines to trustworthy literature values \cite{Fowler2017,Rodriguez2016,Mauron2003}. 
This method relies on the data and estimated uncertainties found in literature and may not always be practical due to missing or interpolated data. Especially when determining absolute energy positions of fluorescence lines it is generally advisable to apply an independent energy calibration. 
Here, the energy calibration of the emission lines is determined by directly referencing the energy scale of the spectrometer to the well-characterized energy scale of the four-crystal-monochromator beamline at the synchrotron radiation facility BESSY II\cite{Krumrey2001}.
This is done by elastic scattering of monochromatized photons with known energy. The excitation energies are tuned to various expected energy positions of the L-emission lines of Gd to ensure numerous data points across all four spectrometer settings.

\begin{figure}
	\begin{center}
	 \def\svgwidth{0.6\columnwidth} 
	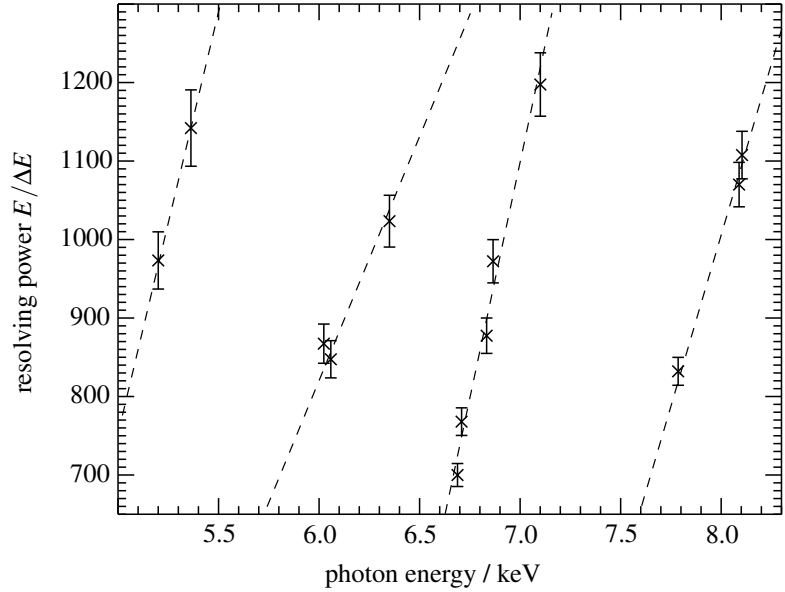
	\end{center}
	\caption{Resolving power \(E/\Delta E\) as a function of photon energy for the four chosen spectrometer settings indicated by the dashed lines. As the photon energy in each setting increases, the radius of the detected circular diffraction patterns decreases and the resolving power increases.}
		\label{pic:resolving_power}
\end{figure}

The distance from sample to CCD highly affects the energy scale of the detected spectrum as shown in Equation (\ref{eq:energy_scale}). To account for different sample positions, i.e., spectrometer source positions (Fig. \ref{pic:von_Hamos} (a)), when switching between the Gd and Au sample, the scattering experiment is repeated on the Gd sample with only one energy per measurement position since scattering on the Gd sample is less efficient. Comparing the radius channel of the scatter peaks from both samples, the derived energy scale from the Au sample is then corrected accordingly for the Gd sample. Corrections are in the order of 1 eV.
In Figure \ref{pic:Gd_energy_axis_pos2} (left) the combined spectrum of the elastically scattered photons from the Au foil for five different photon energies is displayed. The peak positions of the scatter peaks are plotted as a function of radius channel in Figure \ref{pic:Gd_energy_axis_pos2} (right). These data points are fitted with the expression given in Equation (\ref{eq:energy_scale}), connecting photon energy \(E\) to radius channel \(r\) on the basis of the spectrometer geometry:
\begin{align}
\begin{aligned}
E(r) = \left \{ \sin \left[ \arctan \left ( \frac{r s_{\text{pixel}} + 2 R_{\text{HAPG}} }{l} \right ) \right ] \frac{2 d}{n h c}\right \}^{-1},
\label{eq:energy_scale}
\end{aligned}  
\end{align} 
where \(s_{\text{pixel}} = 13.5\,\mu m\) corresponds to the pixel size, \(d=0.3354\) nm to the HAPG lattice constant\cite{Malzer2018}, \(n = 1\) to the Bragg order, \(h\) to Planck’s constant and \(c\) to the speed of light in vacuum. Fit parameters are the distance \(l\) of the CCD to the sample, as well as the effective radius \(R_{\text{HAPG}}\) of the HAPG optics. Equation (\ref{eq:energy_scale}) is essentially derived from the conversion from circle radius to Bragg angle given in Anklamm et al.\cite{Anklamm2014}. Due to a radius dependent response function the scatter peaks slightly increase in width with increasing radius (decreasing photon energy)\cite{Anklamm2014}. The scatter peaks are fitted with a convolution of a Gaussian, representing the energy distribution of the excitation beam, and a Lorentzian, corresponding to the response function. Although it has been reported that the response function is asymmetric with respect to the radius\cite{Anklamm2014}, the symmetric Lorentzian was chosen as best and stable fit due to the statistics of the scatter peaks and the comparatively low spectral resolution.
The deconvolution of the Gd L-emission spectra is then performed on the basis of these response functions assuming a Lorentzian as the natural shape of the emission lines. Figure \ref{pic:resolving_power} displays the resolving power of the spectrometer derived from the individual scatter peaks. The resolving power is defined as \(E/\Delta E\) with \(\Delta E\) being the FWHM of the response function.

\cleardoublepage
\section{Results} 

\begin{figure}
	\begin{center}
	 \def\svgwidth{1.0\columnwidth} 
	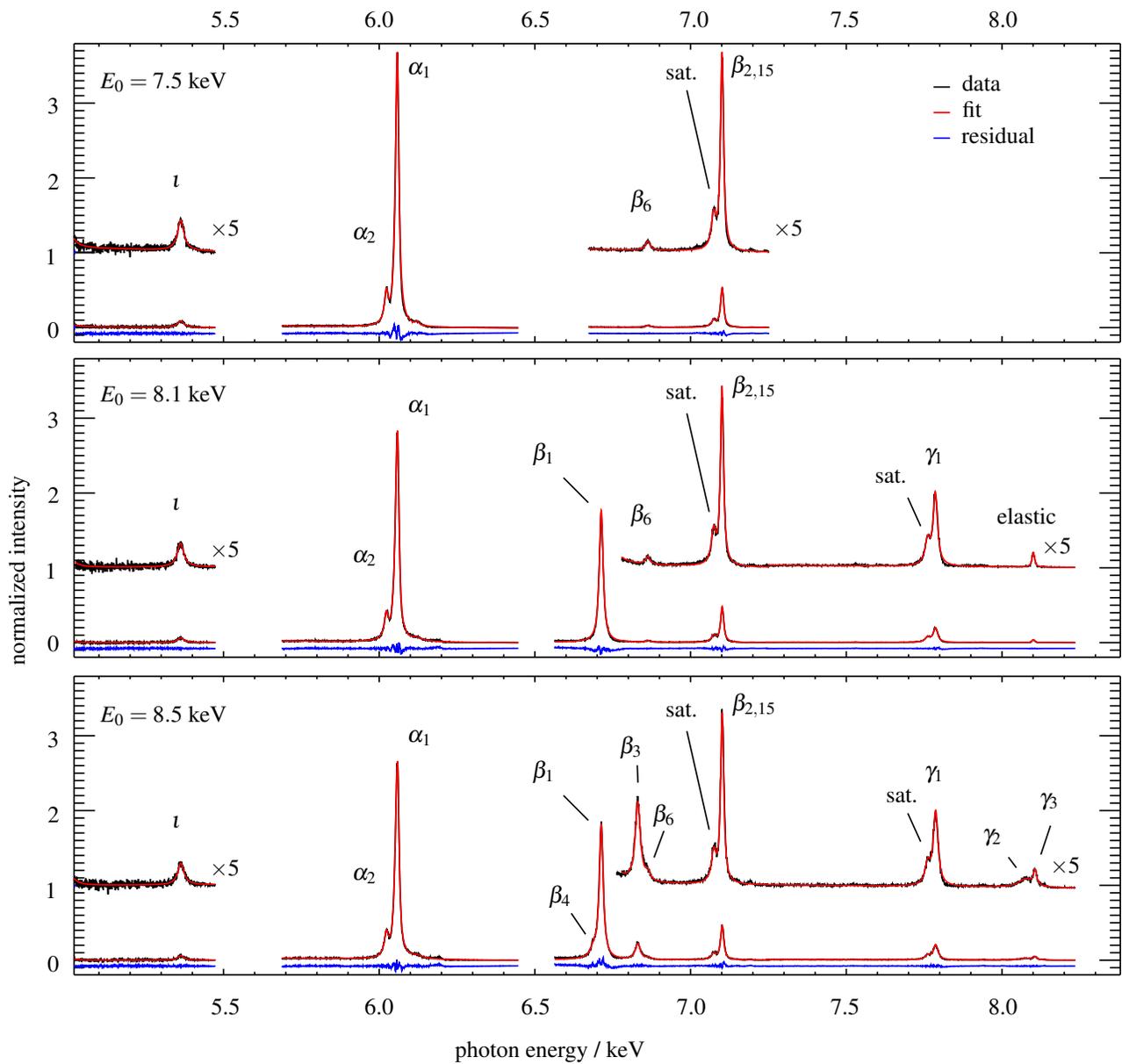
	\end{center}
	\caption{Gd L-emission spectra with excitation of the L$_3$-subshell (7.5 keV), L$_3$- and L$_2$-subshells (8.1 keV), and L$_3$-, L$_2$- and L$_1$-subshell (8.5 keV). The data (black) is fitted with a convolution of Lorentzians and the respective spectrometer response (red). The residual (blue) is shifted by -0.1, the insets by +1.0 and multiplied by a factor of 5 for better visibility. The interruptions of the spectra are caused by the different spectrometer settings.}
		\label{pic:Gd_Lspec}
\end{figure}

Figure \ref{pic:Gd_Lspec} displays the complete L-emission spectrum with respective incident photon energies between the subshell absorption edges. The spectra are normalized to exposure time and number of CCD images as well as the incident photon flux. The sequential emerging of additional emission lines is clearly visible as the incident photon energy is tuned over the L2- and L1-edge. 
To account for self-attenuation\cite{Unterumsberger2018a}, the L-edges of Gd were measured in transmission. In order to separate the absorption of the Gd from the sample substrate and cap layer, samples with the substrate only and the substrate with the Al cap layer were measured as well. 


\subsection{Line energies}
\begin{figure}
	\begin{center}
	\def\svgwidth{0.6\columnwidth} 
	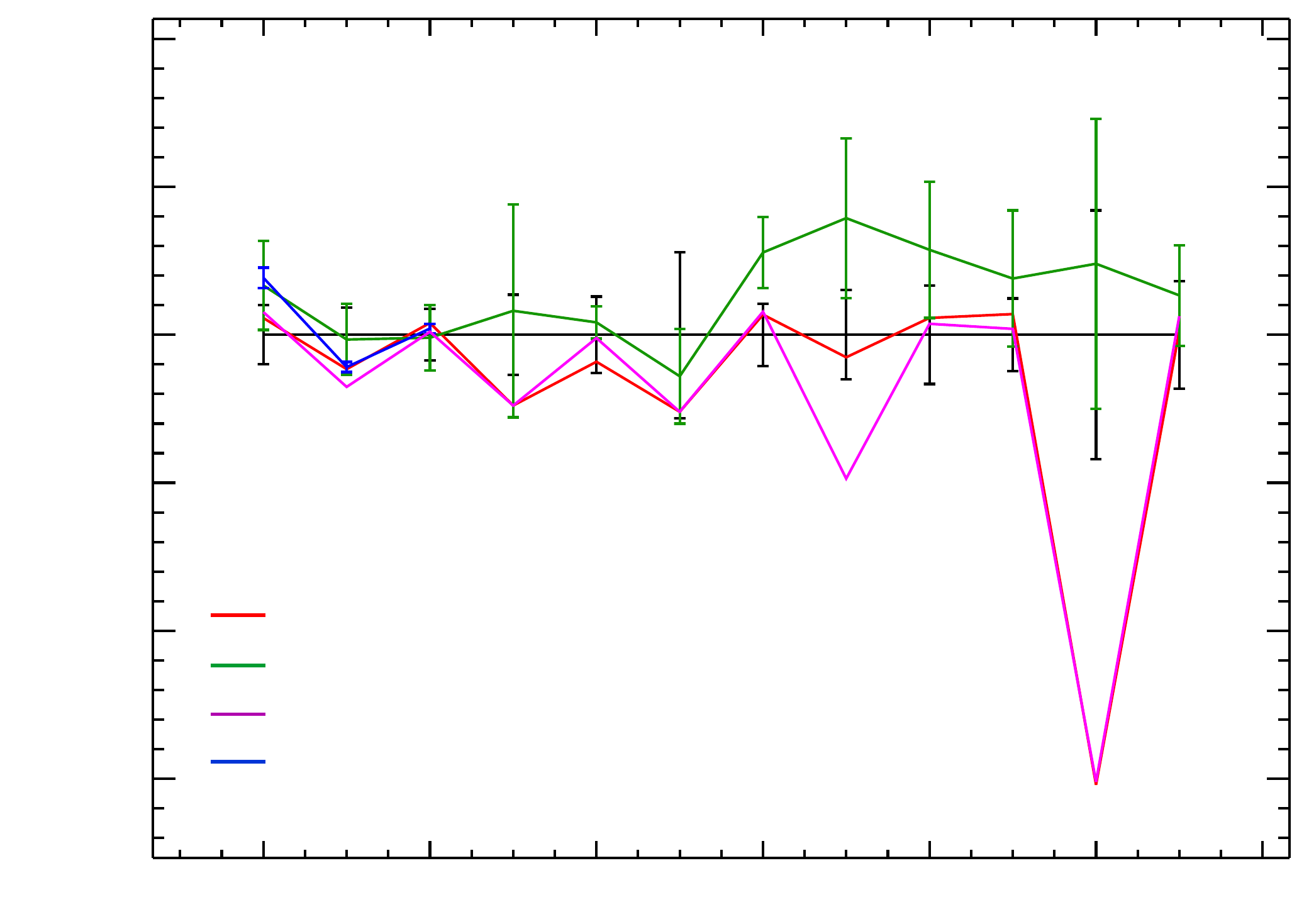
	\end{center} 
	\caption{Visualization of the deviation between experimentally determined Gd L-line energies and published values from Bearden\cite{Bearden1967}, Deslattes\cite{Deslattes2003}, xraylib\cite{Schoonjans2011} and Mauron\cite{Mauron2003}. The values are taken from Table \ref{tab:Gd_L_Energien}.}
	\label{pic:Gd_Line_Energies}
\end{figure}

\begin{figure}
	\begin{center}
	 \def\svgwidth{0.6\columnwidth} 
	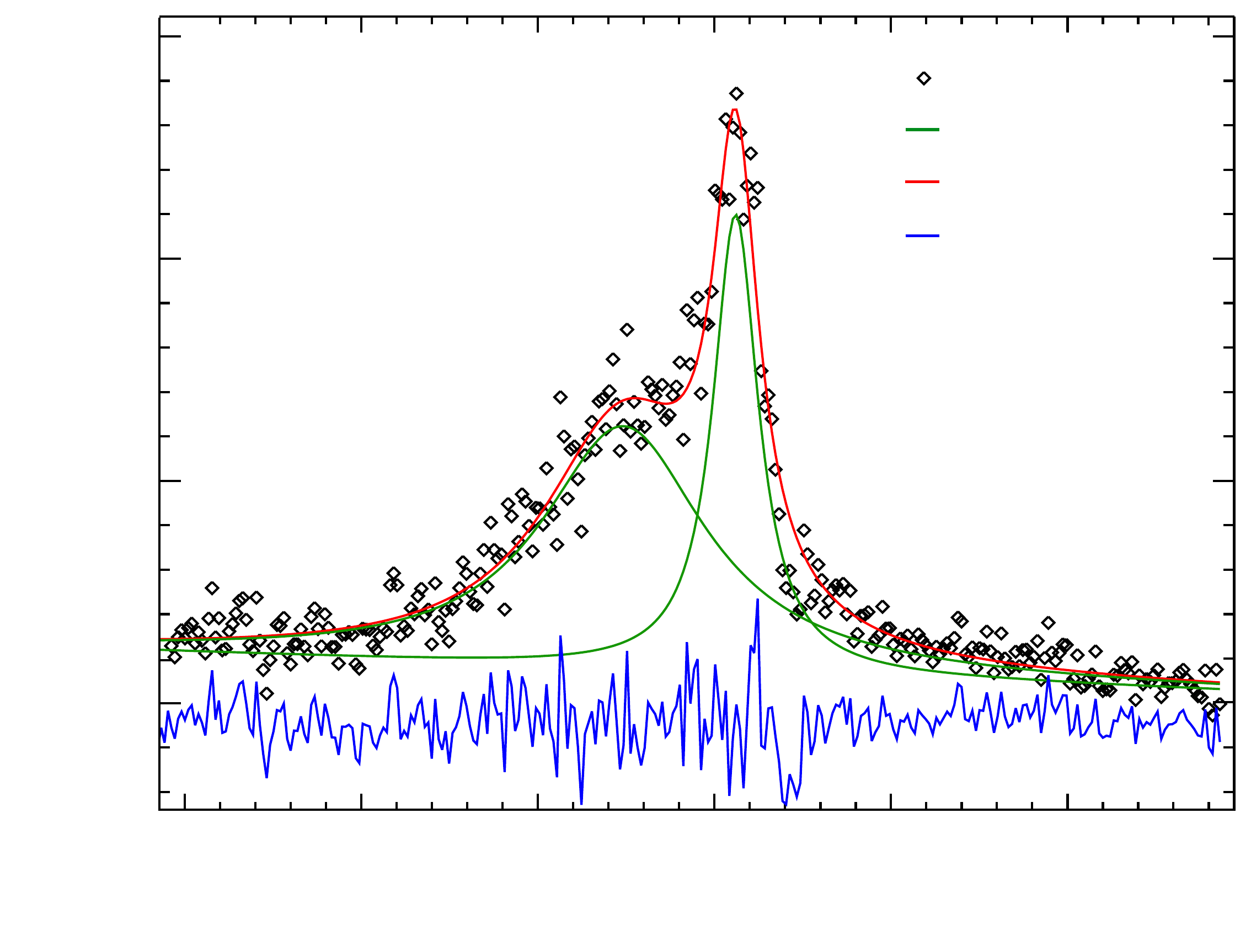
	\end{center}
	\caption{Emission spectrum of the Gd L\(\gamma_{2}\) and L\(\gamma_{3}\) line with incident photon energy of 8.5 keV. The residual (blue) is shifted to -0.1 for better visibility.}
		\label{pic:Gd_Lgamma23}
\end{figure}

In Table \ref{tab:Gd_L_Energien} the experimentally determined Gd L-line energies of this work are compared with values found in databases\cite{Deslattes2003,Bearden1967,Schoonjans2011} and a publication by Mauron et al.\cite{Mauron2003}. The given experimental uncertainties consist of three independent contributions: the beamline resolution from which the energy scale of the spectrometer is derived (0.5 eV to 0.8 eV absolute uncertainty), the polar integration uncertainty which involves finding the correct center of the detected circles and the finite pixel size (0.2 eV to 0.5 eV absolute uncertainty), and the uncertainty of the nonlinear least squares fitting procedure (0.1 eV to 2.9 eV absolute uncertainty). The uncertainty of the fitting procedure is estimated by the standard deviation of the fit results of the spectra obtained from the different excitation energies.
For the sake of intuition the deviation of the experimental values from the literature values is plotted in Figure \ref{pic:Gd_Line_Energies}. The best overall agreement within the given error bars, except for the L\(\gamma_2\) line, is observed for values from Bearden. The Deslattes values tend to be slightly lower in energy, however, agree within the combined uncertainties. Very good agreement with all four references is achieved for the L\(\alpha_1\) line. This is not surprising because the L\(\alpha_1\) line exhibits the highest intensity in the L-emission spectrum. The strongest disagreement is observed in the L\(\gamma_2\) line energy. This emission line is experimentally observed as a very broad low-energy shoulder of the L\(\gamma_3\) line as depicted in Figure \ref{pic:Gd_Lgamma23}. The mismatch of 15.0 eV between the Bearden and xraylib value and our experimental value is strikingly large. The Deslattes value for the L\(\gamma_2\) line, however, deviates only by 2.4 eV from our experimental value, agreeing within the given uncertainty and, hence, supporting our experimental result. 

\begin{figure}
	\begin{center}
	 \def\svgwidth{0.6\columnwidth} 
	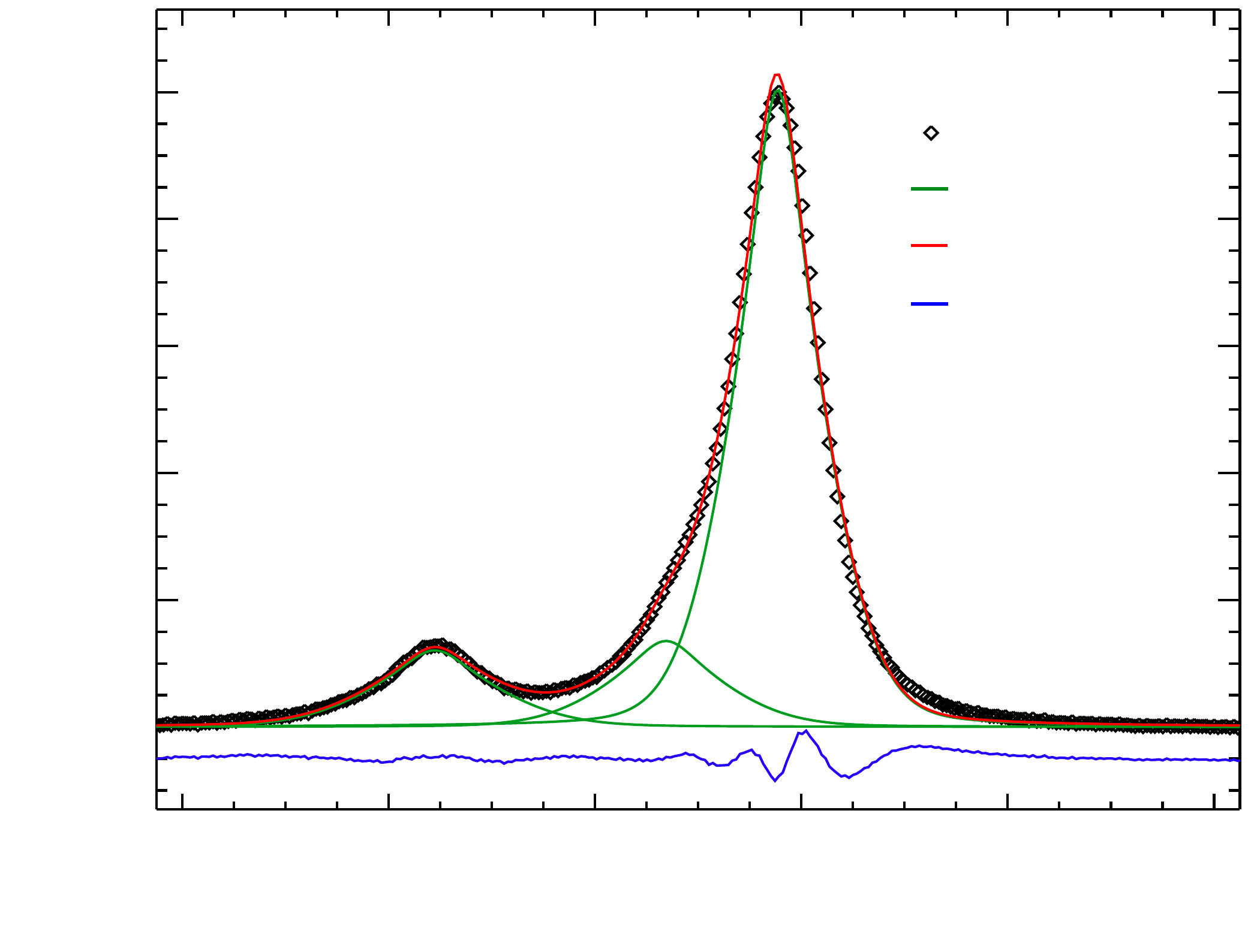
	\end{center}
	\caption{Emission spectrum of Gd L$\alpha_{1,2}$ and L$\eta$ normalized to the maximum of L$\alpha_{1}$. This spectrum is measured with the spectrometer in the double crystal configuration to increase the resolving power. The polychromatic dipole white light synchrotron radiation beamline served as excitation source\cite{Thornagel2001}. The residual (blue) is shifted by -0.1 for better visibility. }
		\label{pic:Gd_eta}
\end{figure}

Two things should be noted: First, the L$\gamma_1$ and L$\beta_{2,15}$ emission lines exhibit satellite lines on the low-energy side which are discussed in section \ref{sec:satellite}. Second, the L\(\eta\) emission line is not observed in the emission spectra. This is owed to the fact that its line energy is expected to be between the L\(\alpha_1\) and L\(\alpha_2\) lines and the achieved experimental resolving power does not allow for resolving this line. As the used von Hamos spectrometer is equipped with a second Bragg crystal enabling higher resolving power, we measured the L\(\alpha_{1,2}\) emission lines again in a double crystal configuration of the spectrometer with a polychromatic excitation source\cite{Thornagel2001} similar to an experiment described in detail elsewhere\cite{Wansleben2018}. The result is displayed in Figure \ref{pic:Gd_eta} revealing the L$\eta$ emission line. Note that the energy scale is derived using the line energies determined for L\(\alpha_{1,2}\).

\begin{table}   
\centering
 \caption{Comparison of the determined line energies to values found in the literature. Good agreement within the given uncertainty is observed for lines with high transition probability (e.g. L\(\alpha_1\), L\(\beta_1\)) and isolated lines (eg. L\(\iota\)). High discrepancies are obtained for close lying emission lines (eg. L\(\gamma_2\) and L\(\gamma_3\)).}
 \begin{tabular}{l l l c c c c c}
 \toprule
subshell	    &	line				    &transition         &  \multicolumn{5}{c}{energy / eV} \\ \midrule
				&  Siegbahn                 & IUPAC             & exp.					& Bearden \cite{Bearden1967}	&Deslattes \cite{Deslattes2003} 	& Mauron \cite{Mauron2003} & xraylib\cite{Schoonjans2011}\\ \midrule
L\(_3\)			&	L\(\iota\)		   	    & L\(_3\)M\(_1\)    &5362.8(9)				& 5362.1						&5361.1(15) 	& 5360.84(34) & 5362.0\\
				&	L\(\alpha_2\)		   	& L\(_3\)M\(_4\)    &6023.8(9)				& 6025.0						&6024.0(12)	    & 6024.93(17) & 6025.6 \\
				&	L\(\alpha_1\)		   	& L\(_3\)M\(_5\)    &6057.7(9)				& 6057.2						&6057.8(11)	    & 6057.49(16) & 6057.6\\
				&	L\(\beta_6\)		    & L\(_3\)N\(_1\)    &6864.6(36)				& 6867.1						&6863.8(36)	    & -           & 6867.0  \\
				&	L\(\beta_{2,15}\)	    & L\(_3\)N\(_{4,5}\)&7102.2(13)				& 7102.8						&7101.77(55)	& -           & 7102.3\\\midrule
L\(_2\)			&   L\(\eta\)               & L\(_2\)M\(_1\)    & 6046.9(28)$^{a}$& 6049.5                        &6048.3(16)     &-            & 6049.5 \\
                &	L\(\beta_1\)            & L\(_2\)M\(_{4}\)  &6713.9(10)				& 6713.2						& 6711.1(12)	& -           & 6713.1 \\
				&	L\(\gamma_1\)		   	& L\(_2\)N\(_{4}\)  &7784.9(15)				& 7785.8						&7781.0(27)	    & -           & 7789.8\\ \midrule
L\(_1\)			&	L\(\beta_4\)		    & L\(_1\)M\(_2\)    &6687.7(15)				& 6687.1						& 6684.8(23)	& -           & 6687.3    \\
				&	L\(\beta_3\)		    & L\(_1\)M\(_3\)    &6831.8(12)				& 6831.1						&6829.9(23)	    & -           & 6831.6\\
				&	L\(\gamma_2\)		    & L\(_1\)N\(_{2}\)  &8072.0(41)				& 8087.0						&8069.6(49)	    & -           & 8087.1 \\
				&	L\(\gamma_3\)		    & L\(_1\)N\(_{3}\)  &8105.3(17)				& 8105.0						&8104.0(17)	    & -           & 8104.7\\ \midrule
				\multicolumn{6}{l}{$^{a}$ determined with double crystal configuration and polychromatic excitation}
 \end{tabular}
 \label{tab:Gd_L_Energien}
 \end{table}

\subsection{Natural line widths and relative transition probabilities \label{sec:widths_TP}}
An accurate determination of the natural line width of emission lines requires foremost the knowledge of the spectral resolution and therewith the spectral response of the spectrometer. As the spectral resolution and response is derived from elastically scattered photons (cf. section \ref{sec:calibration}), the spectra are deconvoluted assuming a Lorentzian as natural line shape. The experimental results and existing references\cite{Perkins1991,Salem1976,Mauron2003} are listed in Table \ref{tab:Gd_L_width}. Good agreement within the combined uncertainty is achieved for the two measured emission lines associated with a vacancy in the L\(_2\) subshell. This is also true for the emission lines associated with a vacancy in the L\(_3\) subshell with the exception of L\(\alpha_{1,2}\). Here, the determined line widths are roughly 30\% higher as compared to literature. The emission lines associated with a vacancy in the L\(_1\) subshell are divided in two groups: The L\(\beta_{3,4}\) line widths agree well within the given uncertainty, the L\(\gamma_{2,3}\) do not. Especially the line width of the L\(\gamma_2\) line shows strong deviations with respect to the references. As Figure \ref{pic:Gd_Lgamma23} suggests, the L\(\gamma_2\) and L\(\gamma_3\) emission lines might have a more complex structure than primordially suggested resulting in the observed deviations.

The experimental uncertainty of the line widths is estimated in the same way as for the line energies. Note that the derived uncertainties of the experimental values are relatively large ( > 1 eV) as compared to the values published by Mauron et al.\cite{Mauron2003} who estimated the spectral response with K\(\alpha_1\) emission lines of 3d elements. The large uncertainties are mainly caused by the moderate resolving power achieved with the single crystal von Hamos spectrometer during the course of the experiments (cf. Figure \ref{pic:resolving_power}).  

\begin{table}
\centering
 \caption{Experimentally determined natural line widths of the Gd L-lines. }
 \begin{tabular}{l l l c c c c}
 \toprule
subshell	       	&  	line   & transition        &  \multicolumn{4}{c}{natural line width / eV} \\ \midrule
                    &  Siegbahn         & IUPAC                    & exp.  & Perkins\cite{Perkins1991} & Salem\cite{Salem1976} & Mauron\cite{Mauron2003} \\ \midrule
 L\(_3\)            & L\(\iota\)       & L\(_3\)M\(_1\)    & 20.5(20)   & 19.36 &  -    & 17.35(65) \\  
                    & L\(\alpha_2\)    & L\(_3\)M\(_4\)    &  7.6(16)   & 4.88  & 4.90  &\\
                    & L\(\alpha_1\)    & L\(_3\)M\(_5\)    &  6.5(13)   & 4.50  & 4.46 & 4.72(45) \\
                    & L\(\beta_6\)     & L\(_3\)N\(_1\)    & 11.8(21)   & 14.19 & - & \\
                    & L\(\beta_{2,15}\)& L\(_3\)N\(_{4,5}\)& 6.1(19)    & 5.55  & 7.70 & \\ \midrule
L\(_2\)             & L\(\eta\)        & L\(_2\)M\(_1\)    & not resolved & 19.25 & - \\
                    & L\(\beta_1\)     & L\(_2\)M\(_{4}\)  &  6.3(18)   & 5.04 & 4.63 & \\
                    & L\(\gamma_1\)    & L\(_2\)N\(_{4}\)  &  7.3(22)   & 5.86  & 7.83 & \\ \midrule
L\(_1\)             & L\(\beta_4\)     & L\(_1\)M\(_2\)    &  9.8(14)   & 12.27 & 9.08 & \\
                    & L\(\beta_3\)     & L\(_1\)M\(_3\)    & 11.6(20)   & 12.71 & 11.20 & \\
                    & L\(\gamma_2\)    & L\(_1\)N\(_{2}\)  & 44.3(74)   & 13.39 & 4.90 & \\
                    & L\(\gamma_3\)    & L\(_1\)N\(_{3}\)  &   5.1(26)  & 11.46 &  -   & \\ \midrule
\end{tabular}
\label{tab:Gd_L_width}
\end{table}
 
\begin{figure}
	\begin{center}
	 \def\svgwidth{0.6\columnwidth} 
	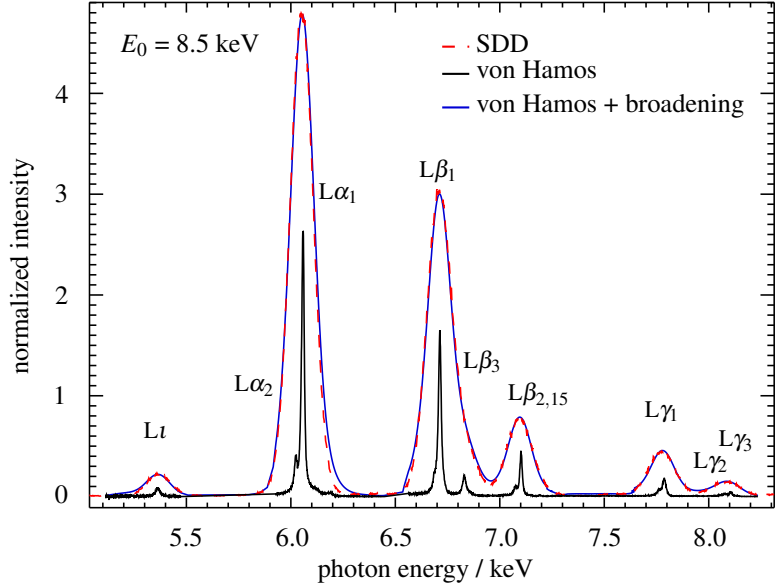
	\end{center}
	\caption{Referencing the high-resolution Gd L-emission spectrum measured with the von Hamos spectrometer to the same measurement with a calibrated SDD. The SDD spectrum and the broadened von Hamos spectrum is scaled for better visibility.}
	\label{pic:Gd_Hamos_SDD}
\end{figure}

The relative transition probabilities refer to the individual subshell line intensity with respect to the total subshell emission intensity. In order to correct for the energy dependent solid angle of detection of the spectrometer, reflectivity of the HAPG crystal, inhomogeneity of the HAPG crystal as well as efficiency of the CCD camera, the measured high-resolution emission spectra are referenced to measurements conducted with a calibrated SDD. This is done by convolution of the measured high-resolution spectra with a Gaussian function which estimates the energy resolution of the SDD resulting in an artificial broadening. The ratio of measured SDD spectrum and broadened high-resolved spectrum of the von Hamos spectrometer results in an energy dependent correction factor. The SDD is calibrated in terms of detection efficiency\cite{Krumrey2006} and the response is experimentally derived and physically modeled\cite{Scholze2009}. A comparison of the Gd L-emission spectra measured with the two spectrometers is displayed in Figure \ref{pic:Gd_Hamos_SDD} where the high-resolved spectrum is already corrected. Both experiments were performed on the same Gd sample using the same set of excitation energies delivered by the FCM synchrotron radiation beamline. Absorption correction is applied using the determined energy dependent mass attenuation of the Gd layer. 
The experimental results and respective values from Elam database\cite{Elam2002} and xraylib\cite{Schoonjans2011} are displayed in Table \ref{tab:Gd_L_TP}. The intensities of the satellite lines of the L\(\gamma_{1}\) and L\(\beta_{2,15}\) line are added to their respective main line. The relative uncertainty of the L$\beta_4$ line is estimated with roughly 30\% as it is energetically very close the L$\beta_1$ emission line. The relative uncertainties of the remaining lines is estimated with 10\% based on the applied referencing procedure and the assumption that the detection efficiency of the von Hamos spectrometer is constant within the energy resolution of the SDD. The results for the emission lines associated with the L\(_3\) and L\(_2\) subshell coincide well with the given values of Elam\cite{Elam2002} and xraylib\cite{Schoonjans2011}. This is, however, not the case for the emission lines associated with the L\(_1\) subshell with the strongest deviations again in the case of the L$\gamma_2$ and L$\gamma_3$ emission lines. While Elam and xraylib predict a more intensive L$\gamma_3$ emission line as compared to the L$\gamma_2$ emission line, our results indicate the opposite. 

\begin{table}
\centering
 \caption{Experimentally determined relative transition probabilities of the Gd L-lines. For L\(\beta_{2,15}\) and L\(\gamma_1\) their respective satellite line is included in the transition probability. }
 \begin{tabular}{l l l c c c }
 \toprule
 subshell	       	&  	line            & transition	    &  \multicolumn{3}{c}{relative transition probabilities} \\ \midrule
                    &  Siegbahn         & IUPAC             & exp.          & Elam\cite{Elam2002} & xraylib\cite{Schoonjans2011} \\ \midrule
 L\(_3\)            & L\(\iota\)        & L\(_3\)M\(_1\)    & 0.033(3)      & 0.030   & 0.034 \\  
                    & L\(\alpha_2\)     & L\(_3\)M\(_4\)    & 0.084(7)      & 0.081   & 0.083 \\
                    & L\(\alpha_1\)     & L\(_3\)M\(_5\)    & 0.745(64)     & 0.728   & 0.733 \\
                    & L\(\beta_6\)      & L\(_3\)N\(_1\)    & 0.009(1)      & 0.007   & 0.008 \\
                    & L\(\beta_{2,15}\) & L\(_3\)N\(_{4,5}\)& 0.129(11)     & 0.153   & 0.139 \\ \midrule
 L\(_2\)            & L\(\eta\)         & L\(_2\)M\(_1\)    & not resolved  & 0.020   & 0.022 \\
                    & L\(\beta_1\)      & L\(_2\)M\(_{4}\)  & 0.846(85)     & 0.839   & 0.825 \\
                    & L\(\gamma_1\)     & L\(_2\)N\(_{4}\)  & 0.154(15)     & 0.141   & 0.144 \\ \midrule
 L\(_1\)            & L\(\beta_4\)      & L\(_1\)M\(_2\)    & 0.176(53)     & 0.293   & 0.312 \\
                    & L\(\beta_3\)      & L\(_1\)M\(_3\)    & 0.567(57)     & 0.483   & 0.446 \\
                    & L\(\gamma_2\)     & L\(_1\)N\(_{2}\)  & 0.166(17)     & 0.093   & 0.074 \\
                    & L\(\gamma_3\)     & L\(_1\)N\(_{3}\)  & 0.091(9)     & 0.131   & 0.110 \\ \midrule
 \end{tabular}
 \label{tab:Gd_L_TP}
 \end{table}

\subsection{ L$\beta_{2,15}$- and L$\gamma_{1}$-satellite \label{sec:satellite}}

\begin{figure}
	\begin{center}
	 \def\svgwidth{0.6\columnwidth} 
	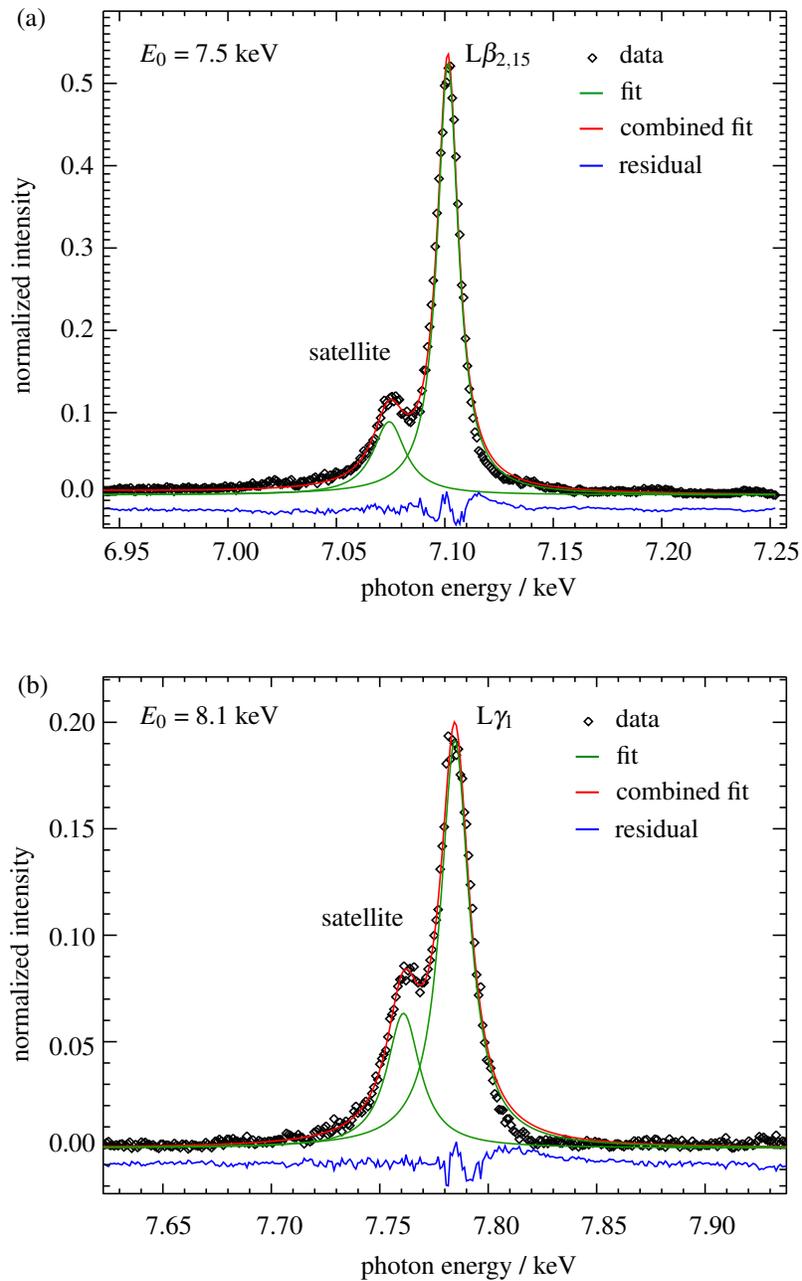
	\end{center}
	\caption{Emission spectrum of (a) Gd L\(\beta_{2,15}\) with an excitation energy of 7.5 keV and (b) of Gd L\(\gamma_{1}\) with an excitation energy of 8.1 keV. The data is fitted with two Lorentzians that are convoluted with the response function of the spectrometer. The residual (blue) is shifted for better visibility.}
		\label{pic:Gd_Lbeta_2_15_L_gamma_1}
\end{figure}

The emission lines L\(\beta_{2,15}\) and L\(\gamma_{1}\) result from electronic transitions from the N\(_{4,5}\)-subshells (4d\(_{5/2}\), 4d\(_{3/2}\)) to the L$_3$ subshell (2p\(_{3/2}\)) and the L$_2$ subshell (2p\(_{1/2}\)). Both of these emission lines exhibit a pronounced shoulder on the low-energy side which is identified as a satellite line (cf. Figure \ref{pic:Gd_Lbeta_2_15_L_gamma_1} (a) and (b)). In general, the term satellite line labels all types of emission lines that are not identified as diagram lines without specifying their physical origin. They are not restricted to either side of the diagram line, however, remain in the energetic vicinity. Contamination with other elements resulting in emission lines that correspond to the energy of the detected satellite lines can be excluded due to mismatching emission energies. Since none of the other diagram lines of the Gd L-emission spectrum originating from M- and N-subshells to the respective L-subshell show satellites within the experimental resolution, it suggests itself to be related to the N\(_{4,5}\)-subshell.

The low-energy satellites of L\(\beta_{2,15}\) and L\(\gamma_{1}\) have been subject to both experimental studies and theoretical studies. Demekhin et al.\cite{Demekhin1972} were the first to conduct experimental studies on these satellites in REM oxides. They observed a correlation of the energy separations of the satellite to the main line and the 4f spin angular momentum and concluded a strong exchange interaction between the final state 4d hole (4d\(^9\)) and 4f electrons as the origin of these satellites. They also found that the intensity ratio of satellite to main line, referred to as branching ratio, is larger in the case of the L\(\gamma_{1}\) line than in the case of the L\(\beta_{2,15}\) line. This is confirmed by our measurements with respective branching ratios of 0.34(2) and 0.27(2). Tanaka et al.\cite{Tanaka1995} developed a theory of XES in REM compounds using an impurity Anderson model. They achieved good agreements with the experimental data of Demekhin et al. by including the multiplet couplings according to the Kramers-Heisenberg formula in the second order optical process. They could also assign the different satellite to main line ratios of L\(\beta_{2,15}\) and L\(\gamma_{1}\) to spin-orbit coupling of the 4d orbitals in the final state and of the 4f orbitals in the initial state. Based on this, another experimental study of REM trifluorides and CeO\(_2\) by Jouda et al.\cite{Jouda1997} could confirm this theoretical framework. 
Studies on elements exceeding the REM elements with completed 4f orbitals such as tungsten (W, Z=74)\cite{Vlaicu1998} and gold (Au, Z=79)\cite{Oohashi2003} show no sign of a low-energy satellite in the case of L\(\beta_{2,15}\), substantiating the previously discussed theory of 4d-4f exchange interaction. 

Table \ref{tab:Gd_sat} summarizes the quantitative analysis with respect to energy spacing and branching ratio of the two satellites in comparison to their main line. The experimental uncertainties are derived according to the description given in sections \ref{sec:calibration} and \ref{sec:widths_TP}. Most of the previous works have characterized these satellites qualitatively. With the established calibration of the spectrometer we could contribute quantitatively to the analysis of these two pronounced satellite lines in the Gd L-emission spectrum. 

\begin{table}
\centering
 \caption{Comparison of the energy spacing and branching ratio (intensity ratio) of L\(\gamma_{1}\) and \(\beta_{2,15}\) to their respective satellite peak.  }
 \begin{tabular}{l l l l l }
 \toprule
    Reference                   &  \multicolumn{2}{c}{\(\Delta E\) / eV}                        & \multicolumn{2}{c}{branching ratio} \\ \midrule 
                                & \(\beta_{2,15}\) - sat.  & L\(\gamma_{1}\) - sat.   & \(I_{sat.}/I_{\beta_{2,15}}\) & \(I_{sat.}/I_{\gamma_{1}}\)\\ \midrule
    Exp.                        & 27.1(25)                      & 23.9(32)                      & 0.27(2) & 0.34(2) \\
    Demekhin\cite{Demekhin1972,Corporation1977} & 25.0                          & 21.0                          & 0.22 & 0.34 \\
    Jouda\cite{Jouda1997}       & 27.1\(^{b}\)                 & 21.9\(^{b}\)                 & -& -\\
    Maddox\cite{Maddox2006}     & -                             & 22.1\(^{b}\)                 & -& -\\ \midrule
    \multicolumn{3}{l}{\(^{b}\)value estimated from figure} 
 \end{tabular}
 \label{tab:Gd_sat}
 \end{table}

\section{Conclusion}
We studied the complete L-emission spectrum of Gd by means of a full-cylinder von Hamos spectrometer and monochromatized synchrotron radiation. Due to the large energy bandwidth of over 600 eV enabled through the spectrometer design, only four spectrometer settings were necessary to cover all L-emission lines. This facilitated the calibration procedure and reduced the time investment necessary to complete measurements. The thorough beamline based calibration of the spectrometer with respect to its energy scale and spectral response allowed for the determination of line energies independently of any reference emission line or database. Deviations of up to 15 eV in the energy of the L\(\gamma_2\) line to the Bearden database\cite{Bearden1967} were found. Additionally, we investigated two low-energy satellites of the L\(\gamma_{1}\) line and the L\(\beta_{2,15}\) line which are not found in common databases\cite{Bearden1967,Deslattes2003} but have been subject to previous research\cite{Demekhin1972,Tanaka1995,Jouda1997,Maddox2006}. On the basis of these existing publications including theory and experiment, we were able to update the quantitative analysis of these two lines with respect to energy spacing and relative intensity ratio (branching ratio). Furthermore, we quantified the natural line widths and relative transition probabilities of the measured L-lines. 

The transition probabilities of emission lines associated with a respective vacancy in the L$_2$ and L$_3$ subshell agree with the references, rather large discrepancies are found in the emission lines associated with a vacancy in the L$_1$ subshell, especially for the L$\gamma_{2,3}$ lines. Similar results are obtained for the natural line widths with strong deviations in the case of the L$\gamma_{2,3}$ lines with respect to the references. Strong deviations in all three FP's, line energy, line widths, and relative transition probabilities, for the L$\gamma_{2,3}$ emission lines are observed. It should be noted that the Deslattes database already suggests that the L$\gamma_2$ emission line is expected to not have a distinct shape due to a complex level structure.
Examining the emission spectrum of L\(\gamma_{2,3}\) with improved statistics and resolution in the measurement would be beneficial to examine and identify respective transitions and investigate additional satellite lines accompanying the diagram transitions. 

A limiting factor contributing to the overall experimental uncertainties is the moderate spectral resolving power of the spectrometer varying between 700 and 1200. In order to increase the resolving power, we see potential in reducing the source size by using a different focusing optics. We further showed, in the case of the L$\eta$ line, that the use of the double crystal concept improves the resolution significantly which enabled us to determine the L$\eta$ line energy. Using two crystals to investigate the full L-emission spectrum, however, requires more energy intervals to be arranged for by corresponding HAPG crystal positionings and related calibrations, as well as longer measurement times to account for a decreased detection efficiency caused by the second Bragg reflection. 

\section{Acknowledgement}
Valuable discussions with J. Vinson, NIST, and Prof. B. Kanngießer, TU Berlin, are most appreciated. This research was performed within the EMPIR project AEROMET. The financial support of the EMPIR program is gratefully acknowledged. It is jointly funded by the European Metrology Programme for Innovation and Research (EMPIR) and participating countries within the European Association of National Metrology Institutes (EURAMET) and the European Union.

\cleardoublepage
\bibliography{bib}

\begin{thebibliography}{10}
\urlstyle{rm}
\expandafter\ifx\csname url\endcsname\relax
  \def\url#1{\texttt{#1}}\fi
\expandafter\ifx\csname urlprefix\endcsname\relax\def\urlprefix{URL }\fi
\expandafter\ifx\csname doiprefix\endcsname\relax\def\doiprefix{DOI: }\fi
\providecommand{\bibinfo}[2]{#2}
\providecommand{\eprint}[2][]{\url{#2}}

\bibitem{Beckhoff2008a}
\bibinfo{author}{Beckhoff, B.}
\newblock \bibinfo{journal}{\bibinfo{title}{{Reference-free X-ray spectrometry
  based on metrology using synchrotron radiation}}}.
\newblock {\emph{\JournalTitle{J. Anal. At. Spectrom.}}}
  \textbf{\bibinfo{volume}{23}}, \bibinfo{pages}{845--853},
  \doiprefix\url{10.1039/b718355k} (\bibinfo{year}{2008}).

\bibitem{Honicke2014}
\bibinfo{author}{H{\"{o}}nicke, P.} \emph{et~al.}
\newblock \bibinfo{journal}{\bibinfo{title}{{Experimental verification of the
  individual energy dependencies of the partial L -shell photoionization cross
  sections of Pd and Mo}}}.
\newblock {\emph{\JournalTitle{Phys. Rev. Lett.}}}
  \textbf{\bibinfo{volume}{113}}, \bibinfo{pages}{163001},
  \doiprefix\url{10.1103/PhysRevLett.113.163001} (\bibinfo{year}{2014}).

\bibitem{Menesguen2015}
\bibinfo{author}{M{\'{e}}nesguen, Y.} \emph{et~al.}
\newblock \bibinfo{journal}{\bibinfo{title}{{High accuracy experimental
  determination of copper and zinc mass attenuation coefficients in the 100 eV
  to 30 keV photon energy range}}}.
\newblock {\emph{\JournalTitle{Metrologia}}} \textbf{\bibinfo{volume}{53}},
  \bibinfo{pages}{7--17}, \doiprefix\url{10.1088/0026-1394/53/1/7}
  (\bibinfo{year}{2015}).

\bibitem{Honicke2016}
\bibinfo{author}{H{\"{o}}nicke, P.}, \bibinfo{author}{Kolbe, M.},
  \bibinfo{author}{Krumrey, M.}, \bibinfo{author}{Unterumsberger, R.} \&
  \bibinfo{author}{Beckhoff, B.}
\newblock \bibinfo{journal}{\bibinfo{title}{{Experimental determination of the
  oxygen K-shell fluorescence yield using thin SiO2 and Al2O3 foils}}}.
\newblock {\emph{\JournalTitle{Spectrochim. Acta Part B At. Spectrosc.}}}
  \textbf{\bibinfo{volume}{124}}, \bibinfo{pages}{94--98},
  \doiprefix\url{https://doi.org/10.1016/j.sab.2016.08.024}
  (\bibinfo{year}{2016}).

\bibitem{Unterumsberger2018}
\bibinfo{author}{Unterumsberger, R.} \emph{et~al.}
\newblock \bibinfo{journal}{\bibinfo{title}{{Accurate experimental
  determination of gallium K- and L3-shell XRF fundamental parameters}}}.
\newblock {\emph{\JournalTitle{J. Anal. At. Spectrom.}}}
  \textbf{\bibinfo{volume}{33}}, \bibinfo{pages}{1003--1013},
  \doiprefix\url{10.1039/c8ja00046h} (\bibinfo{year}{2018}).

\bibitem{Holfelder2018}
\bibinfo{author}{Holfelder, I.} \emph{et~al.}
\newblock \bibinfo{journal}{\bibinfo{title}{{A Compact and Calibratable von
  Hamos X-Ray Spectrometer Based on Two Full-Cylinder HAPG Mosaic Crystals for
  High-Resolution XES}}}.
\newblock {\emph{\JournalTitle{Proc. 10th Mech. Eng. Des. Synchrotron Radiat.
  Equip. Instrum.}}} \bibinfo{pages}{189--192},
  \doiprefix\url{doi:10.18429/JACoW-MEDSI2018-WEOPMA06} (\bibinfo{year}{2018}).

\bibitem{Tokumitsu2000}
\bibinfo{author}{Tokumitsu, H.} \emph{et~al.}
\newblock \bibinfo{journal}{\bibinfo{title}{{Gadolinium neutron-capture therapy
  using novel gadopentetic acid- chitosan complex nanoparticles: In vivo growth
  suppression of experimental melanoma solid tumor}}}.
\newblock {\emph{\JournalTitle{Cancer Lett.}}} \textbf{\bibinfo{volume}{150}},
  \bibinfo{pages}{177--182}, \doiprefix\url{10.1016/S0304-3835(99)00388-2}
  (\bibinfo{year}{2000}).

\bibitem{Caravan1999}
\bibinfo{author}{Caravan, P.}, \bibinfo{author}{Ellison, J.~J.},
  \bibinfo{author}{McMurry, T.~J.} \& \bibinfo{author}{Lauffer, R.~B.}
\newblock \bibinfo{journal}{\bibinfo{title}{{Gadolinium(III) chelates as MRI
  contrast agents: Structure, dynamics, and applications}}}.
\newblock {\emph{\JournalTitle{Chem. Rev.}}} \textbf{\bibinfo{volume}{99}},
  \bibinfo{pages}{2293--2352}, \doiprefix\url{10.1021/cr980440x}
  (\bibinfo{year}{1999}).

\bibitem{Carley2012}
\bibinfo{author}{Carley, R.} \emph{et~al.}
\newblock \bibinfo{journal}{\bibinfo{title}{{Femtosecond laser excitation
  drives ferromagnetic gadolinium out of magnetic equilibrium}}}.
\newblock {\emph{\JournalTitle{Phys. Rev. Lett.}}}
  \textbf{\bibinfo{volume}{109}}, \bibinfo{pages}{057401},
  \doiprefix\url{10.1103/PhysRevLett.109.057401} (\bibinfo{year}{2012}).

\bibitem{Fowler2017}
\bibinfo{author}{Fowler, J.~W.} \emph{et~al.}
\newblock \bibinfo{journal}{\bibinfo{title}{{A reassessment of absolute
  energies of the x-ray L lines of lanthanide metals}}}.
\newblock {\emph{\JournalTitle{Metrologia}}} \textbf{\bibinfo{volume}{54}},
  \bibinfo{pages}{494--511}, \doiprefix\url{10.1088/1681-7575/aa722f}
  (\bibinfo{year}{2017}).
\newblock \eprint{1702.00507}.

\bibitem{Krumrey1998}
\bibinfo{author}{Krumrey, M.}
\newblock \bibinfo{journal}{\bibinfo{title}{{Design of a four-crystal
  monochromator beamline for radiometry at BESSY II}}}.
\newblock {\emph{\JournalTitle{J. Synchrotron Radiat.}}}
  \textbf{\bibinfo{volume}{5}}, \bibinfo{pages}{6--9},
  \doiprefix\url{10.1107/S0909049597011825} (\bibinfo{year}{1998}).

\bibitem{Krumrey2001}
\bibinfo{author}{Krumrey, M.} \& \bibinfo{author}{Ulm, G.}
\newblock \bibinfo{journal}{\bibinfo{title}{{High-accuracy detector calibration
  at the PTB four-crystal monochromator beamline}}}.
\newblock {\emph{\JournalTitle{Nucl. Instruments Methods Phys. Res. Sect. A
  Accel. Spectrometers, Detect. Assoc. Equip.}}}
  \textbf{\bibinfo{volume}{467-468}}, \bibinfo{pages}{1175--1178},
  \doiprefix\url{10.1016/S0168-9002(01)00598-8} (\bibinfo{year}{2001}).

\bibitem{Gerlach2008}
\bibinfo{author}{Gerlach, M.} \emph{et~al.}
\newblock \bibinfo{journal}{\bibinfo{title}{{Cryogenic radiometry in the hard
  x-ray range}}}.
\newblock {\emph{\JournalTitle{Metrologia}}} \textbf{\bibinfo{volume}{45}},
  \bibinfo{pages}{577--585}, \doiprefix\url{10.1088/0026-1394/45/5/012}
  (\bibinfo{year}{2008}).

\bibitem{Kolbe2012}
\bibinfo{author}{Kolbe, M.}, \bibinfo{author}{H{\"{o}}nicke, P.},
  \bibinfo{author}{M{\"{u}}ller, M.} \& \bibinfo{author}{Beckhoff, B.}
\newblock \bibinfo{journal}{\bibinfo{title}{{L-subshell fluorescence yields and
  Coster-Kronig transition probabilities with a reliable uncertainty budget for
  selected high- and medium-Z elements}}}.
\newblock {\emph{\JournalTitle{Phys. Rev. A - At. Mol. Opt. Phys.}}}
  \textbf{\bibinfo{volume}{86}}, \bibinfo{pages}{042512},
  \doiprefix\url{10.1103/PhysRevA.86.042512} (\bibinfo{year}{2012}).

\bibitem{Anklamm2014}
\bibinfo{author}{Anklamm, L.} \emph{et~al.}
\newblock \bibinfo{journal}{\bibinfo{title}{{A novel von Hamos spectrometer for
  efficient X-ray emission spectroscopy in the laboratory}}}.
\newblock {\emph{\JournalTitle{Rev. Sci. Instrum.}}}
  \textbf{\bibinfo{volume}{85}}, \bibinfo{pages}{053110},
  \doiprefix\url{10.1063/1.4875986} (\bibinfo{year}{2014}).

\bibitem{Malzer2018}
\bibinfo{author}{Malzer, W.} \emph{et~al.}
\newblock \bibinfo{journal}{\bibinfo{title}{{A laboratory spectrometer for high
  throughput X-ray emission spectroscopy in catalysis research}}}.
\newblock {\emph{\JournalTitle{Rev. Sci. Instrum.}}}
  \textbf{\bibinfo{volume}{89}}, \bibinfo{pages}{113111},
  \doiprefix\url{10.1063/1.5035171} (\bibinfo{year}{2018}).

\bibitem{Wansleben2018}
\bibinfo{author}{Wansleben, M.}, \bibinfo{author}{Vinson, J.},
  \bibinfo{author}{Holfelder, I.}, \bibinfo{author}{Kayser, Y.} \&
  \bibinfo{author}{Beckhoff, B.}
\newblock \bibinfo{journal}{\bibinfo{title}{{Valence-to-core X-ray emission
  spectroscopy of Ti, TiO, and TiO2 by means of a double full-cylinder crystal
  von Hamos spectrometer}}}.
\newblock {\emph{\JournalTitle{X-Ray Spectrom.}}}
  \textbf{\bibinfo{volume}{48}}, \bibinfo{pages}{102--106},
  \doiprefix\url{10.1002/xrs.3000} (\bibinfo{year}{2018}).

\bibitem{Gerlach2015}
\bibinfo{author}{Gerlach, M.} \emph{et~al.}
\newblock \bibinfo{journal}{\bibinfo{title}{{Characterization of HAPG mosaic
  crystals using synchrotron radiation}}}.
\newblock {\emph{\JournalTitle{J. Appl. Crystallogr.}}}
  \textbf{\bibinfo{volume}{48}}, \bibinfo{pages}{1381--1390},
  \doiprefix\url{10.1107/S160057671501287X} (\bibinfo{year}{2015}).

\bibitem{Schlesiger2017}
\bibinfo{author}{Schlesiger, C.}, \bibinfo{author}{Anklamm, L.},
  \bibinfo{author}{Malzer, W.}, \bibinfo{author}{Gnewkow, R.} \&
  \bibinfo{author}{Kanngie{\ss}er, B.}
\newblock \bibinfo{journal}{\bibinfo{title}{{A new model for the description of
  X-ray diffraction from mosaic crystals for ray-tracing calculations}}}.
\newblock {\emph{\JournalTitle{J. Appl. Crystallogr.}}}
  \textbf{\bibinfo{volume}{50}}, \bibinfo{pages}{1490--1497},
  \doiprefix\url{10.1107/S1600576717012626} (\bibinfo{year}{2017}).

\bibitem{Chabot1991}
\bibinfo{author}{Chabot, M.} \emph{et~al.}
\newblock \bibinfo{journal}{\bibinfo{title}{{X-ray reflectivities, at low and
  high order of reflection, of flat highly oriented pyrolytic graphite
  crystals}}}.
\newblock {\emph{\JournalTitle{Nucl. Inst. Methods Phys. Res. B}}}
  \textbf{\bibinfo{volume}{61}}, \bibinfo{pages}{377--384},
  \doiprefix\url{10.1016/0168-583X(91)95309-2} (\bibinfo{year}{1991}).

\bibitem{Arkadiev2007}
\bibinfo{author}{Arkadiev, V.~A.} \emph{et~al.}
\newblock \bibinfo{journal}{\bibinfo{title}{{X-ray analysis with a highly
  oriented pyrolytic graphite-based von Hamos spectrometer}}}.
\newblock {\emph{\JournalTitle{Spectrochim. Acta Part B At. Spectrosc.}}}
  \textbf{\bibinfo{volume}{62}}, \bibinfo{pages}{577--585},
  \doiprefix\url{10.1016/j.sab.2007.03.003} (\bibinfo{year}{2007}).

\bibitem{Doppner2008}
\bibinfo{author}{D{\"{o}}ppner, T.} \emph{et~al.}
\newblock \bibinfo{journal}{\bibinfo{title}{{High order reflectivity of highly
  oriented pyrolytic graphite crystals for x-ray energies up to 22 keV}}}.
\newblock {\emph{\JournalTitle{Rev. Sci. Instrum.}}}
  \textbf{\bibinfo{volume}{79}}, \bibinfo{pages}{10E311},
  \doiprefix\url{10.1063/1.2966378} (\bibinfo{year}{2008}).

\bibitem{Grigorieva2019}
\bibinfo{author}{Grigorieva, I.}, \bibinfo{author}{Antonov, A.} \&
  \bibinfo{author}{Gudi, G.}
\newblock \bibinfo{journal}{\bibinfo{title}{{Graphite Optics - Current
  Opportunities , Properties and Limits}}}.
\newblock {\emph{\JournalTitle{Condens. Matter}}} \textbf{\bibinfo{volume}{4}},
  \doiprefix\url{10.3390/condmat4010018} (\bibinfo{year}{2019}).

\bibitem{Smolek2010}
\bibinfo{author}{Smolek, S.}, \bibinfo{author}{Streli, C.},
  \bibinfo{author}{Zoeger, N.} \& \bibinfo{author}{Wobrauschek, P.}
\newblock \bibinfo{journal}{\bibinfo{title}{{Improved micro x-ray fluorescence
  spectrometer for light element analysis}}}.
\newblock {\emph{\JournalTitle{Rev. Sci. Instrum.}}}
  \textbf{\bibinfo{volume}{81}}, \bibinfo{pages}{053707},
  \doiprefix\url{10.1063/1.3428739} (\bibinfo{year}{2010}).

\bibitem{Kayser2014}
\bibinfo{author}{Kayser, Y.} \emph{et~al.}
\newblock \bibinfo{journal}{\bibinfo{title}{{Laboratory-based micro-X-ray
  fluorescence setup using a von Hamos crystal spectrometer and a focused beam
  X-ray tube}}}.
\newblock {\emph{\JournalTitle{Rev. Sci. Instrum.}}}
  \textbf{\bibinfo{volume}{85}}, \bibinfo{pages}{043101},
  \doiprefix\url{10.1063/1.4869340} (\bibinfo{year}{2014}).

\bibitem{Unterumsberger2012}
\bibinfo{author}{Unterumsberger, R.} \emph{et~al.}
\newblock \bibinfo{journal}{\bibinfo{title}{{Focusing of soft X-ray radiation
  and characterization of the beam profile enabling X-ray emission spectrometry
  at nanolayered specimens}}}.
\newblock {\emph{\JournalTitle{Spectrochim. Acta - Part B At. Spectrosc.}}}
  \textbf{\bibinfo{volume}{78}}, \bibinfo{pages}{37--41},
  \doiprefix\url{10.1016/j.sab.2012.10.001} (\bibinfo{year}{2012}).

\bibitem{Rodriguez2016}
\bibinfo{author}{Rodr{\'{i}}guez, T.}, \bibinfo{author}{Sep{\'{u}}lveda, A.},
  \bibinfo{author}{Carreras, A.}, \bibinfo{author}{Castellano, G.} \&
  \bibinfo{author}{Trincavelli, J.}
\newblock \bibinfo{journal}{\bibinfo{title}{{Structure of the Ru, Ag and Te L
  X-ray emission spectra}}}.
\newblock {\emph{\JournalTitle{J. Anal. At. Spectrom.}}}
  \textbf{\bibinfo{volume}{31}}, \bibinfo{pages}{780--789},
  \doiprefix\url{10.1039/c5ja00498e} (\bibinfo{year}{2016}).

\bibitem{Mauron2003}
\bibinfo{author}{Mauron, O.} \emph{et~al.}
\newblock \bibinfo{journal}{\bibinfo{title}{{Reexamination of L3 and M1
  atomic-level widths of elements 54{\textless}Z{\textless}77}}}.
\newblock {\emph{\JournalTitle{Phys. Rev. A - At. Mol. Opt. Phys.}}}
  \textbf{\bibinfo{volume}{67}}, \bibinfo{pages}{032506},
  \doiprefix\url{10.1103/PhysRevA.67.032506} (\bibinfo{year}{2003}).

\bibitem{Unterumsberger2018a}
\bibinfo{author}{Unterumsberger, R.}, \bibinfo{author}{H{\"{o}}nicke, P.},
  \bibinfo{author}{Pollakowski-Herrmann, B.}, \bibinfo{author}{M{\"{u}}ller,
  M.} \& \bibinfo{author}{Beckhoff, B.}
\newblock \bibinfo{journal}{\bibinfo{title}{{Relative L3 transition
  probabilities of titanium compounds as a function of the oxidation state
  using high-resolution X-ray emission spectrometry}}}.
\newblock {\emph{\JournalTitle{Spectrochim. Acta Part B At. Spectrosc.}}}
  \textbf{\bibinfo{volume}{145}}, \bibinfo{pages}{71--78},
  \doiprefix\url{10.1016/j.sab.2018.04.008} (\bibinfo{year}{2018}).

\bibitem{Deslattes2003}
\bibinfo{author}{Deslattes, R.~D.} \emph{et~al.}
\newblock \bibinfo{journal}{\bibinfo{title}{{X-ray transition energies: New
  approach to a comprehensive evaluation}}}.
\newblock {\emph{\JournalTitle{Rev. Mod. Phys.}}}
  \textbf{\bibinfo{volume}{75}}, \bibinfo{pages}{35--99},
  \doiprefix\url{10.1103/RevModPhys.75.35} (\bibinfo{year}{2003}).

\bibitem{Bearden1967}
\bibinfo{author}{Bearden, J.~A.}
\newblock \bibinfo{journal}{\bibinfo{title}{{X-Ray Wavelengths}}}.
\newblock {\emph{\JournalTitle{Rev. Mod. Phys.}}}
  \textbf{\bibinfo{volume}{39}}, \bibinfo{pages}{78--124},
  \doiprefix\url{10.1103/RevModPhys.39.78} (\bibinfo{year}{1967}).

\bibitem{Schoonjans2011}
\bibinfo{author}{Schoonjans, T.} \emph{et~al.}
\newblock \bibinfo{journal}{\bibinfo{title}{{The xraylib library for
  X-ray-matter interactions. Recent developments}}}.
\newblock {\emph{\JournalTitle{Spectrochim. Acta Part B At. Spectrosc.}}}
  \textbf{\bibinfo{volume}{66}}, \bibinfo{pages}{776--784},
  \doiprefix\url{10.1016/j.sab.2011.09.011} (\bibinfo{year}{2011}).

\bibitem{Thornagel2001}
\bibinfo{author}{Thornagel, R.}, \bibinfo{author}{Klein, R.} \&
  \bibinfo{author}{Ulm, G.}
\newblock \bibinfo{journal}{\bibinfo{title}{{The electron storage ring BESSY II
  as a primary source standard from the visible to the the X-ray range}}}.
\newblock {\emph{\JournalTitle{Metrologia}}} \textbf{\bibinfo{volume}{38}},
  \bibinfo{pages}{385--389}, \doiprefix\url{10.1088/0026-1394/38/5/3}
  (\bibinfo{year}{2001}).

\bibitem{Perkins1991}
\bibinfo{author}{Perkins, S.} \emph{et~al.}
\newblock \bibinfo{title}{{Tables and graphs of atomic subshell and relaxation
  data derived from the LLNL Evaluated Atomic Data Library (EADL), Z =
  1--100}}.
\newblock \bibinfo{type}{Tech. Rep.}, \bibinfo{institution}{Lawrence Livermore
  National Lab., CA (United States)} (\bibinfo{year}{1991}).
\newblock \doiprefix\url{10.2172/10121422}.

\bibitem{Salem1976}
\bibinfo{author}{Salem, S.~I.} \& \bibinfo{author}{Lee, P.~L.}
\newblock \bibinfo{journal}{\bibinfo{title}{{Experimental widths of K and L
  x-ray lines}}}.
\newblock {\emph{\JournalTitle{At. Data Nucl. Data Tables}}}
  \textbf{\bibinfo{volume}{18}}, \bibinfo{pages}{233--241},
  \doiprefix\url{10.1016/0092-640X(76)90026-7} (\bibinfo{year}{1976}).

\bibitem{Krumrey2006}
\bibinfo{author}{Krumrey, M.}, \bibinfo{author}{Gerlach, M.},
  \bibinfo{author}{Scholze, F.} \& \bibinfo{author}{Ulm, G.}
\newblock \bibinfo{journal}{\bibinfo{title}{{Calibration and characterization
  of semiconductor X-ray detectors with synchrotron radiation}}}.
\newblock {\emph{\JournalTitle{Nucl. Instruments Methods Phys. Res. Sect. A
  Accel. Spectrometers, Detect. Assoc. Equip.}}}
  \textbf{\bibinfo{volume}{568}}, \bibinfo{pages}{364--368},
  \doiprefix\url{10.1016/j.nima.2006.06.004} (\bibinfo{year}{2006}).

\bibitem{Scholze2009}
\bibinfo{author}{Scholze, F.} \& \bibinfo{author}{Procop, M.}
\newblock \bibinfo{journal}{\bibinfo{title}{{Modelling the response function of
  energy dispersive X-ray spectrometers with silicon detectors}}}.
\newblock {\emph{\JournalTitle{X-Ray Spectrom.}}}
  \textbf{\bibinfo{volume}{38}}, \bibinfo{pages}{312--321},
  \doiprefix\url{10.1002/xrs.1165} (\bibinfo{year}{2009}).

\bibitem{Elam2002}
\bibinfo{author}{Elam, W.~T.}, \bibinfo{author}{Ravel, B.~D.} \&
  \bibinfo{author}{Sieber, J.~R.}
\newblock \bibinfo{journal}{\bibinfo{title}{{A new atomic database for X-ray
  spectroscopic calculations}}}.
\newblock {\emph{\JournalTitle{Radiat. Phys. Chem.}}}
  \textbf{\bibinfo{volume}{63}}, \bibinfo{pages}{121--128},
  \doiprefix\url{10.1016/S0969-806X(01)00227-4} (\bibinfo{year}{2002}).

\bibitem{Demekhin1972}
\bibinfo{author}{Demekhin, V.~F.}, \bibinfo{author}{Platkov, A.~I.} \&
  \bibinfo{author}{Lyubivaya, M.~V.}
\newblock \bibinfo{journal}{\bibinfo{title}{{Multiplet Structure Investigation
  of the L-Series Lines of Rare Earth Elements}}}.
\newblock {\emph{\JournalTitle{Sov. Phys. JETP}}}
  \textbf{\bibinfo{volume}{35}}, \bibinfo{pages}{28--31}
  (\bibinfo{year}{1972}).

\bibitem{Tanaka1995}
\bibinfo{author}{Tanaka, S.}, \bibinfo{author}{Ogasawara, H.},
  \bibinfo{author}{Okada, K.} \& \bibinfo{author}{Kotani, A.}
\newblock \bibinfo{journal}{\bibinfo{title}{{Theory of the 4d -{\textgreater}2p
  X-Ray Emission Spectroscopy in Ce2 O3, Pr2O3 and Dy2 O3}}}.
\newblock {\emph{\JournalTitle{J. Phys. Soc. Japan}}}
  \textbf{\bibinfo{volume}{64}}, \bibinfo{pages}{2225--2232},
  \doiprefix\url{10.1143/JPSJ.64.2225} (\bibinfo{year}{1995}).

\bibitem{Jouda1997}
\bibinfo{author}{Jouda, K.}, \bibinfo{author}{Tanaka, S.} \&
  \bibinfo{author}{Aita, O.}
\newblock \bibinfo{journal}{\bibinfo{title}{{L$\gamma$1 and L$\beta$2,15 x-ray
  emission lines of rare-earth trifluorides and CeO2}}}.
\newblock {\emph{\JournalTitle{J. Phys. Condens. Matter}}}
  \textbf{\bibinfo{volume}{9}}, \bibinfo{pages}{10789--10794},
  \doiprefix\url{10.1088/0953-8984/9/48/020} (\bibinfo{year}{1997}).

\bibitem{Vlaicu1998}
\bibinfo{author}{Vlaicu, A.-M.} \emph{et~al.}
\newblock \bibinfo{journal}{\bibinfo{title}{{Investigation of the 74W L
  emission spectra and satellites}}}.
\newblock {\emph{\JournalTitle{Phys. Rev. A - At. Mol. Opt. Phys. At. Mol. Opt.
  Phys.}}} \textbf{\bibinfo{volume}{58}}, \bibinfo{pages}{3544--3551},
  \doiprefix\url{10.1103/PhysRevA.58.3544} (\bibinfo{year}{1998}).

\bibitem{Oohashi2003}
\bibinfo{author}{Oohashi, H.}, \bibinfo{author}{Tochio, T.},
  \bibinfo{author}{Ito, Y.} \& \bibinfo{author}{Vlaicu, A.~M.}
\newblock \bibinfo{journal}{\bibinfo{title}{{Origin of Au L$\beta$2 visible
  satellites}}}.
\newblock {\emph{\JournalTitle{Phys. Rev. A - At. Mol. Opt. Phys.}}}
  \textbf{\bibinfo{volume}{68}}, \bibinfo{pages}{032506},
  \doiprefix\url{10.1103/PhysRevA.68.032506} (\bibinfo{year}{2003}).

\bibitem{Corporation1977}
\bibinfo{author}{Demekhin, V.~F.}, \bibinfo{author}{Yavna, S.~A.},
  \bibinfo{author}{Bairachnyi, Y.} \& \bibinfo{author}{Sukhorukov, V.~L.}
\newblock \bibinfo{journal}{\bibinfo{title}{{The one-configuration
  approximation in the calculation of the x-ray and electron spectra of the
  transition elements}}}.
\newblock {\emph{\JournalTitle{J. Struct. Chem.}}}
  \textbf{\bibinfo{volume}{18}}, \bibinfo{pages}{513--519}
  (\bibinfo{year}{1977}).

\bibitem{Maddox2006}
\bibinfo{author}{Maddox, B.~R.} \emph{et~al.}
\newblock \bibinfo{journal}{\bibinfo{title}{{4f delocalization in Gd: Inelastic
  X-ray scattering at ultrahigh pressure}}}.
\newblock {\emph{\JournalTitle{Phys. Rev. Lett.}}}
  \textbf{\bibinfo{volume}{96}}, \bibinfo{pages}{215701},
  \doiprefix\url{10.1103/PhysRevLett.96.215701} (\bibinfo{year}{2006}).

\end{thebibliography}

\end{document}